\documentclass[fleqn,usenatbib,useAMS]{mnras}

\usepackage[T1]{fontenc}

\DeclareRobustCommand{\VAN}[3]{#2}
\let\VANthebibliography\thebibliography
\def\thebibliography{\DeclareRobustCommand{\VAN}[3]{##3}\VANthebibliography}

\usepackage{graphicx}	% Including figure files
\usepackage{amssymb}	% Extra maths symbols
\usepackage{color}
\usepackage{enumitem}
\usepackage{hyperref}
\usepackage{verbatim}
\usepackage{amsmath,amstext,mathrsfs}
\usepackage[all]{hypcap} 
\usepackage{afterpage}    
\usepackage{marginnote}
\usepackage{makecell}
\usepackage{subfigure}
\usepackage{multirow}
\usepackage{siunitx}

\title[Multi-scale Magnetic Fields in the Central Molecular Zone]{Multi-scale Magnetic Fields in the Central Molecular Zone: Inference from the Gradient Technique}
\author[Hu et al.]{
Yue Hu$^{1,2}$\thanks{E-mail: yue.hu@wisc.edu}
,A. Lazarian$^{2,3}$\thanks{E-mail: alazarian@facstaff.wisc.edu}
,Q. Daniel Wang$^{4}$\thanks{E-mail: wqd@astro.umass.eduu}
\\
$^{1}$Department of Physics, University of Wisconsin-Madison, Madison, WI, 53706, USA\\
$^{2}$Department of Astronomy, University of Wisconsin-Madison, Madison, WI, 53706, USA\\
$^{3}$Centro de Investigación en Astronomía, Universidad Bernardo O’Higgins, Santiago, General Gana 1760, 8370993,
Chile\\
$^{4}$Department of Astronomy, University of Massachusetts, Amherst, MA 01003, USA\\
}

\date{Accepted XXX. Received YYY; in original form ZZZ}

\pubyear{2020}

\begin{document}
\label{firstpage}
\pagerange{\pageref{firstpage}--\pageref{lastpage}}
\maketitle

% Abstract of the paper
\begin{abstract}
The central molecular zone (CMZ)  plays an essential role in regulating the nuclear ecosystem of our Galaxy. To get an insight into magnetic fields of the CMZ, we employ the Gradient Technique (GT), which is rooted in the anisotropy of magnetohydrodynamic turbulence. Our analysis is based on the data of multiple wavelengths, including molecular emission lines, radio 1.4 GHz continuum image, and Herschel \SI{70}{\micro\meter} image, as well as ionized [Ne II] and Paschen-alpha emissions. The results are compared with the observations of Planck 353 GHz and High-resolution Airborne Wideband Camera Plus (HWAC+) \SI{53}{\micro\meter} polarized dust emissions. We map the magnetic fields orientation at multiple wavelength across the central molecular zone, including  close-ups of the Radio Arc and Sagittarius A West regions, on multi scales from $\sim$ 0.1 pc to 10 pc. The magnetic fields towards the central molecular zone traced by GT are globally compatible with the polarization measurements, accounting for the contribution from the galactic foreground and background. This correspondence suggests that the magnetic field and turbulence are dynamically crucial in the galactic center. We find that the magnetic fields associated with the Arched filaments and the thermal components of the Radio Arc are in good agree with the HAWC+ polarization. Our measurement towards the non-thermal Radio Arc reveals the poloidal magnetic field components in the galactic center. For Sagittarius A West region, we find a great agreement between the GT measurement using [Ne II] emission and HWAC+ \SI{53}{\micro\meter} observation. We use GT to predict the magnetic fields associated with ionized Paschen-alpha gas down to scales of 0.1 pc.
%These results demonstrate the potential power of GT in high-resolution mapping of magnetic fields and in decomposing contributions from different velocity components and/or different gas phases.
\end{abstract}

% Select between one and six entries from the list of approved keywords.
% Don't make up new ones.

%%%%%%%%%%%%%%%%%%%%%%%%%%%%%%%%%%%%%%%%%%%%%%%%%%

%%%%%%%%%%%%%%%%% BODY OF PAPER %%%%%%%%%%%%%%%%%%
\section{Introduction}
\label{sec:intro}
The central molecular zone (CMZ), which is a vast reservoir of dense molecular gas, dominates the interstellar medium (ISM) around the galactic center \citep{1996ARA&A..34..645M,1998ApJS..118..455O,2019PASJ...71S..19T}. Despite its intense concentration of dense gas, the estimated star formation rate is approximately one order of magnitude lower than expectation \citep{2013MNRAS.429..987L,2015ApJ...799...53K,2017MNRAS.469.2263B}. Possible explanations include an important role of magnetic fields and turbulence in the galactic center \citep{2010Natur.463...65C,2014MNRAS.440.3370K}, which can regulate the star-formation process \citep{2012ApJ...761..156F,2014PhR...539...49K,2018ApJ...863..118B,Hu20}. To understand this discrepancy, it is essential to study the magnetic field, although it is notoriously challenging. 

In the past decades, several measurements of the interstellar magnetic field in the CMZ have been achieved, including radio, far-infrared, and submillimeter observation \citep{1988ApJ...333..729W,1990ApJ...362..114H,2000ApJ...529..241N,2005MNRAS.360.1305R,2008A&A...478..435R,2007SPIE.6678E..0DV}. The detected non-thermal radio structures, which are shaped as thin and long filaments, reveals a poloidal component of the magnetic field, i.e., perpendicular to the Galactic plane \citep{1986AJ.....92..818T,1990IAUS..140..361M,1999ApJ...521L..41L}. As a follow-up, a large-scale toroidal component was observed by \cite{2003ApJ...583L..83N} through polarized dust emission at \SI{450}{\micro\meter}. Also, the recent Planck survey of polarized dust emission at a wavelength of \SI{850}{\micro\meter} offers a broader view of the large-scale magnetic field morphology in the CMZ \citep{2015A&A...576A.104P}. This large-scale picture is further completed by the PILOT balloon-borne experiment at \SI{240}{\micro\meter} \citep{2019A&A...630A..74M}.

In addition to these measurements, a novel insight into the interstellar magnetic fields can be gain from the application of the Gradient Technique (GT; \citealt{2017ApJ...835...41G,YL17a, LY18a, PCA}). Unlike the synchrotron or dust polarization, GT explores the anisotropy of magnetohydrodynamic (MHD) turbulence, or the statistical elongation of  turbulent eddies along their local magnetic fields that percolate them \citep{GS95,LV99}. It also suggests that at a given separation, the minimum amplitude of velocity fluctuations appears at the direction parallel to local magnetic fields, while the maximum amplitude appears at the perpendicular direction. Consequently, the gradient of velocity fluctuations' amplitude is perpendicular to the local magnetic field. Therefore, similar to the case of polarization, the velocity gradient rotated by 90$^\circ$ reveals the orientation of the magnetic field. The applicability of GT in molecular clouds has been observationally tested by \cite{survey,IRAM}, \cite{2020arXiv200715344A}, and \cite{2021arXiv210913670L}. A similar statistical description  also holds for density fields, which are passively regulated by velocity. Consequently, the density gradient is also perpendicular to the local magnetic field \citep{YL17b, IGs,cluster}. 

Recent high resolution $\rm ^{12}CO$ (1–0), $\rm ^{13}CO$ (1–0), and $\rm HNC$ (1–0) emission lines observed with the Nobeyama 45m telescope \citep{2018ApJS..236...40T,2019PASJ...71S..19T}, now enable us to resolve the CMZ region down to the scale of $~0.5$ pc (FWHM$\approx15''$). At this scale, the role of turbulence is expected to be important. In this context, the GT can be  used to gain valuable information regarding the structure of the magnetic field and the molecular gas density in the CMZ. Furthermore, disentangling the magnetic fields across several wavebands can be realized by the GT. Employing a broad range of wavelengths, our work brings a unique insight to study separately the magnetic fields embedded in different gas phases along the line-of-sight (LOS) in the CMZ. 

In order to have a more comprehensive picture of the magnetic fields, our analysis includes the magnetic fields in neighboring physically different regions. For instance, the Radio Arc, the Arched Filaments, and Sagittarius A West \citep{2010ApJS..191..275L,2012ApJ...755...90I}. The first two structures observed at 1.4 GHz suggest either poloidal or toroidal magnetic fields in the CMZ. While the Sgr A West is observed with [Ne II] emission and it has a very distinctive and extreme physical condition due to the central supermassive black hole. These data, including the molecular lines, the Planck 353 GHz \citep{2020A&A...641A...3P} and HWAC+ \SI{53}{\micro\meter} \citep{2018JAI.....740008H} dust polarization data, allow us to map  the magnetic field at suitable multi-wavelength in both diffuse-ionized-gas region and dense-cold-gas region in multi scales from the order of 10 pc to 0.1 pc. Thus, this work provides a multi-wavelength and multi-scale view of the magnetic fields and their interactions with the multi-phase gas in different parts of the galactic center ecosystem.

The paper is organized as follows. In \S~\ref{sec:data}, we provide the details of the observational data used in this work. In \S~\ref{sec:method}, we illustrate the recipe of GT to trace the magnetic fields. In \S~\ref{sec:results}, we present the magnetic fields traced by GT at multi-wavelength and multi-scale. We make compare the magnetic fields traced by GT with Planck 353 GHz and HAWC+ \SI{53}{\micro\meter} polarization. We give discussion in \S~\ref{sec.dis} and summary in \S~\ref{sec:con}, respectively.

%%%%%%%%%%%%%%%%%%%%%%%%%%%%%%%%%%%%%%%%%%%%%%%%%%%%%
\section{Observational Data}
\label{sec:data}
\subsection{Central Molecular Zone}
The $\rm ^{12}CO$ (1–0), $\rm ^{13}CO$ (1–0), HNC (1-0) emission lines towards the Galactic CMZ were observed with the Nobeyama 45m telescope \citep{2018ApJS..236...40T,2019PASJ...71S..19T}. The data cover the area: $-0.8^\circ <l<1.2^\circ$ and $-0.35^\circ <b<+0.35^\circ$ with a beamwidth of $\rm FWHM\approx15''$ and velocity resolution of $\approx1.3$ km/s for $\rm ^{12}CO$ (1–0) and $\approx0.67$ $km$ $s^{-1}$ for $\rm ^{13}CO$ (1–0). The final data cubes were resampled onto a $7.5'' \times 7.5''\times 2$ km/s and $10.275'' \times 10.275''\times 2$ km/s grid for $\rm ^{12}CO$/$\rm ^{13}CO$ and HNC, respectively. The rms noises of $\rm ^{12}CO$ (1–0), $\rm ^{13}CO$ (1–0), and HNC (1-0) are approximately 1.00 K, 0.20 K, and 0.21 K, respectively. We select the emission within the radial velocity range of -220 to +220 km/s for our analysis.
 	
The CMZ is also observed by the Herschel Space Observatory. The far-infrared Herschel \SI{70}{\micro\meter} image from the Hi-GAL survey \citep{2010PASP..122..314M} is observed with the PACS detector. The FWHM beam size of the PACS spectrometer is $9.2''$ near \SI{70}{\micro\meter}. The pixel size of \SI{70}{\micro\meter} image has been set to $3''$. The 1$\sigma$ sensitivity is 17.6 mJy in the PACS \SI{70}{\micro\meter} band.

\subsection{Radio Arc and Arched Filaments}
The Radio Arc, which is one of the most prominent radio continua features in the Galactic center \citep{1984Natur.310..557Y}, is observed at frequency 1420.406 MHz with the Very Large Array (VLA) telescope \citep{2010ApJS..191..275L}. It achieves a resolution of $\rm FWHM\approx15''$ (0.6 pc) and a total bandwidth of 1.5625 MHz. The RMS noise for the spectra is approximately $1\sigma\approx10$ mJy. 

\subsection{Sagittarius A West}
The [Ne II] emission line for Sgr A West was observed with the Texas Echelon Cross Echelle Spectrograph. At the \SI{12.8}{\micro\meter} wavelength, the [Ne II] finestructure line has a spectral resolution of $\approx 4$ km/s, and a spatial resolution of $\approx1.3''$ \citep{2012ApJ...755...90I}. The  rms noise level is around 0.01 K.

The Sgr A is also observed by hydrogen Paschen-$\alpha$ line (wavelength $\sim\SI{1.876}{\micro\meter}$), using the NICMOS instrument aboard the Hubble Space Telescope \citep{2010MNRAS.402..895W}. It covers the Sgr A in the 1.87 and \SI{1.90}{\micro\meter} narrow bands. Its spatial resolution is around 0.01 pc (FWHM $\sim0.2''$) at a distance of 8 kpc. The rms noise level is around 0.06 mJy arcsec$^{-2}$ for the full resolution image with a pixel size of 0.1$''\times$ 0.1$''$.

\subsection{Polarized dust emission}
The POS magnetic field orientation in the CMZ was inferred from the Planck 353 GHz and HAWC+ polarized dust signal data.

In this work, we use the Planck 3rd Public Data Release (DR3) 2018 of High Frequency Instrument \citep{2020A&A...641A...3P}\footnote{Based on observations obtained with Planck (\url{http://www.esa.int/Planck}), an ESA science mission with instruments and contributions directly funded by ESA Member States, NASA, and Canada.}. The Planck observations defines the polarization angle $\phi$ and polarization fraction $p$ through Stokes parameter maps $I$, $Q$, and $U$:
\begin{equation}
\begin{aligned}
   \phi&=\frac{1}{2}\arctan(-U,Q)\\
   p&=\sqrt{Q^2+U^2}/I
\end{aligned}
\end{equation}
where $-U$ converts the angle from HEALPix convention to IAU convention and the two-argument function $\arctan$ is used to account for the $\pi$ periodicity. To increase the signal-to-noise ratio, we smooth all maps from nominal angular resolution $5'$ up to a resolution of $10'$ using a Gaussian kernel. The magnetic field angle is inferred from $\phi_B=\phi+\pi/2$. The Planck polarization is used to reveal the magnetic fields of the large-scale entire CMZ.

For small-scale magnetic field tracing, we use the HAWC+ polarization measurement obtained from HAWC+ archival database \citep{2018JAI.....740008H}. We select the band A measurement ($\rm FWHM\approx5''$) and only pixels with $p/\sigma_p>3$ are considered, where $p$ is the polarization fraction, and $\sigma_p$ is its uncertainty. Similar to Planck, magnetic field is perpendicular to polarization angle, i.e., $\phi_B=\phi+\pi/2$. We use the HWAC+ polarization to reveal the magnetic field towards the Sickle (i.e., the radio arc), and Sgr A*.

\section{Methodology}
\label{sec:method}
\subsection{Anisotropy of MHD turbulence}
The technique of tracing magnetic fields through velocity gradients is developed from the advanced MHD turbulence theory (see \citealt{GS95}, hereafter GS95) and turbulent reconnection theory (see \citealt{LV99}, hereafter LV99). In what follows, we briefly explain those essentials. 

The prophetic study of GS95 proposed that the turbulent eddy is anisotropic, i.e., the eddy is elongating along magnetic fields. To derive this anisotropy, GS95 considers the "critical balance" condition, i.e., the cascading time ($k_\bot v_l$)$^{-1}$ equals the wave periods ($k_\parallel v_A$)$^{-1}$, and  Kolmogorov-type turbulence, i.e., $v_l\propto l^{1/3}$. Here $k_\parallel$ and $k_\bot$ are wavevectors parallel and perpendicular to the magnetic field, respectively. $v_l$ is turbulent velocity at scale $l$ and $v_A$ is Alfv\'{e}n speed. The corresponding GS95 anisotropy scaling be can easily be obtained: 
\begin{equation}
    k_\parallel\propto k_\bot^{2/3}
\end{equation}
which reveals the anisotropy increases as the scale of turbulent motions decreases. However, this derivation is drawn in Fourier space, which means the direction of wavevectors is defined with respect to the mean magnetic field in the global reference frame. In this global reference frame, small eddies' contribution is averaged out by large eddies. Consequently, one can only observe a scale-independent anisotropy, which are dominated by the largest eddy \citep{2000ApJ...539..273C}.

The anisotropic property of sub-Alf\'enic MHD turluence is later boosted by LV99. LV99 explained that the scale-dependent anisotropy is only observable in the local reference frame, which is defined in real space with respect to the magnetic field passing through the eddy at scale $l$. The consideration employs the notion of fast turbulent reconnection. As local magnetic fields gives minimal resistance along the direction perpendicular to the magnetic field, it is easier to mix the magnetic field lines instead of bending it. Consequently, the turbulent cascading is channeled to the direction perpendicular to the magnetic field, i.e., the motion of eddies perpendicular to the local magnetic field direction obeys the Kolmogorov law $v_{l,\bot}\propto l_\bot^{1/3}$. Here $l_\bot$ and $v_{l,\bot}$ are the perpendicular components of eddies' scale and velocity with respect to the {\it local} magnetic field, respectively. Considering the "critical balance" in the local reference frame: $v_{l,\bot}l_\bot^{-1}\approx v_Al_\parallel^{-1}$, one can obtain the scale-dependent anisotropy scaling:
\begin{equation}
\label{eq.lv99}
 l_\parallel= L_{\rm inj}(\frac{l_\bot}{L_{\rm inj}})^{\frac{2}{3}}{M_A^{-4/3}}, { M_A\le 1}
\end{equation}
where $l_\|$ is the parallel component of eddies' scale. $L_{\rm inj}$ is the turbulence injection scale. This scale-dependent anisotropy in the local reference frame was numerically demonstrated
\citep{2000ApJ...539..273C,2001ApJ...554.1175M} and in-situ observations in solar wind \citep{2016ApJ...816...15W}. In particular, the corresponding anisotropy scaling for velocity fluctuation and gradient of velocity fluctuation are :
\begin{equation}
\label{eq.v}
\begin{aligned}
 v_{l,\bot}&= v_{\rm inj}(\frac{l_\bot}{L_{\rm inj}})^{\frac{1}{3}}{M_A^{1/3}}\\
      \nabla v_l&\propto\frac{v_{l,\bot}}{l_\bot}\simeq \frac{v_{\rm inj}}{L_{\rm inj}}(\frac{l_{\perp}}{L_{\rm inj}})^{-\frac{2}{3}}M_A^{\frac{1}{3}}
\end{aligned}
\end{equation}
where $v_{\rm inj}$ is the injection velocity \citep{LV99}. 

Gradients of velocities, magnetic field, and intensity can arise both from MHD turbulence and from various astrophysical structures. The distinction between the two cases is that in MHD turbulence the gradients are perpendicular to the magnetic field, which may not be the case for other settings. For instance, shocks can produce density gradients that tend to be parallel to magnetic field \citep{2021arXiv211104759X}. Therefore the observed correspondence of gradients and polarization supports the idea of MHD turbulence being the origin of the observed gradients. A very important property of gradients induced by turbulence is the increase of gradient amplitude with the decrease of the scale, which is not usually true for large scale gradients of non-turbulent nature. In our studies. While tracing the magnetic field by gradients is thoroughly tested with an extensive set of numerical simulations \citep{LY18a,IGs,2021ApJ...911...53H}, this does not exclude that in given observations gradients cannot arise due to other reasons. In the paper, we mention alternative processes that can induce gradients for particular settings.  

\begin{figure*}[p]
	\centering
	\includegraphics[width=1.0\linewidth,height=1.12\linewidth]{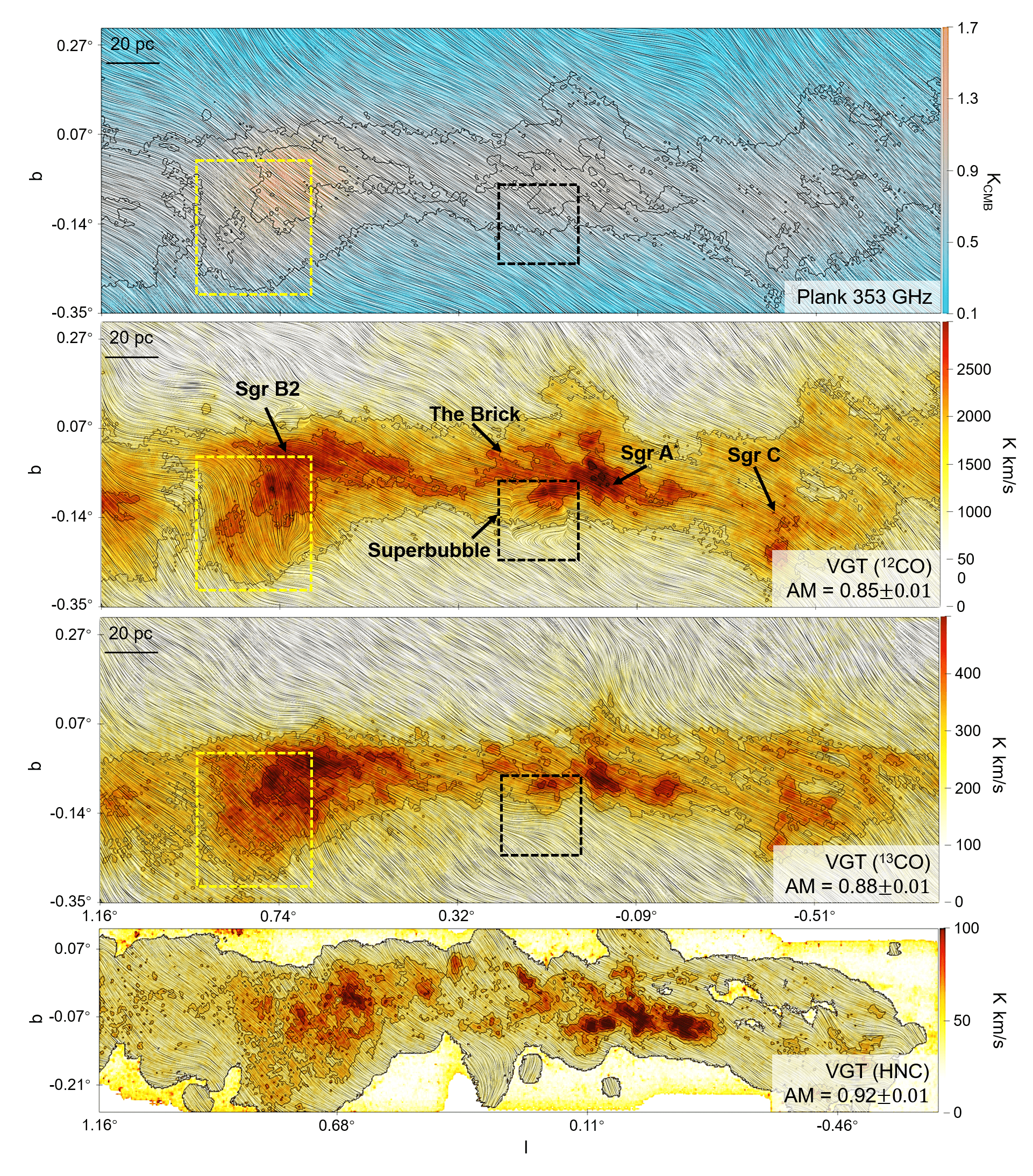}
	\caption{\label{fig:CMZ_B}Visualization of magnetic fields towards the CMZ 		using the Line Integral Convolution (LIC), which is a technique proposed by Cabral \& Leedom (1993) to visualize a vector field. The magnetic fields were inferred from Planck 353 GHz polarized dust emission (top) and the VGT using $\rm ^{12}CO$, $\rm ^{13}CO$, and HNC emissions. The foreground contribution has been subtracted from the Planck polarization and is overlaid on its total dust emission intensity map. The VGT measurements are overlaid on corresponding integrated intensity color maps. The contours on $\rm ^{12}CO$ and $\rm ^{13}CO$ intensity maps start from twice the mean intensity. The $\rm ^{12}CO$'s contours are also overlaid on the Planck map. The yellow and black boxes outlines two regions in which significantly different magnetic field morphology can be seen between the Planck and the $\rm ^{12}CO$ VGT measurements. }
\end{figure*}
\subsection{The Velocity Gradient Technique}
The Velocity Gradient Technique (VGT; \citealt{2017ApJ...835...41G,YL17a,LY18a,PCA}) is the main analysis tool in the work. It is theoretically rooted in the advanced magnetohydrodynamic (MHD) turbulence theory and fast turbulent reconnection theory, as discussed above.
Here we employ thin velocity channel maps \textbf{Ch(x,y)} to extract velocity information in PPV cubes\footnote{The statistics of the intensity fluctuations in PPV and their relations to the underlying statistics of turbulent velocity and density are presented in \cite{LP00}. There it was shown the velocity fluctuations are most prominent in thin channel maps. The criterion for distinguishing the thin channel and the thick channel is given as:
\begin{equation}
\label{eq1}
\begin{aligned}
    \Delta v&<\sqrt{\delta (v^2)}, \thickspace\mbox{thin channel}\\
    \Delta v&\ge\sqrt{\delta (v^2)},\thickspace\mbox{thick channel}\\
\end{aligned}
\end{equation}
where $\Delta v$ is the velocity channel width, $\sqrt{\delta (v^2)}$ is the velocity dispersion corresponding to eddies under study.}.
Each thin channel map is convolved with 3 $\times$ 3 Sobel kernels\footnote{$$
	G_x=\begin{pmatrix} 
	-1 & 0 & +1 \\
	-2 & 0 & +2 \\
	-1 & 0 & +1
	\end{pmatrix} \quad,\quad
	G_y=\begin{pmatrix} 
	-1 & -2 & -1 \\
	0 & 0 & 0 \\
	+1 & +2 & +1
	\end{pmatrix}
	$$}
$G_x$ and $G_y$ to calculate pixelized gradient map $\psi_{g}^i(x,y)$:
\begin{equation}
\label{eq:conv}
\begin{aligned}
\nabla_x Ch_i(x,y)&=G_x * Ch_i(x,y)\\ 
\nabla_y Ch_i(x,y)&=G_y * Ch_i(x,y)\\
\psi_{g}^i(x,y)&=\tan^{-1}\left(\frac{\nabla_y Ch_i(x,y)}{\nabla_x Ch_i(x,y)}\right),
\end{aligned}
\end{equation}
where $\nabla_x Ch_i(x,y)$ and $\nabla_y Ch_i(x,y)$ are the $x$ and $y$ components of gradient, respectively, and $*$ denotes the convolution. To suppresses the effect noise in the spectroscopic data, we mask out the raw gradient whose corresponding intensity is less than three times of RMS noise.

As the orthogonal relative orientation between velocity gradients and the magnetic field appears only when the sampling is statistically sufficient, each raw gradient map $\psi_{g}^i(x,y)$ is further processed by the sub-block averaging method \citep{YL17a}. The sub-block averaging method takes all gradients orientation within a sub-block of interest and then plots the corresponding histogram. A Gaussian fitting is then applied to the histogram. The Gaussian distribution's expectation value is the statistically most probable gradient's orientation within that sub-block.

By repeating the gradient's calculation and the sub-block averaging method for each thin velocity channel (totally $n_v$ thin channel maps), we obtain totally $n_v$ processed gradient maps $\psi_{gs}^i(x,y)$ with $i=1,2,...,n_v$. In analogy to the Stokes parameters of polarization, the pseudo Q$_g$ and U$_g$ of gradient-inferred magnetic fields are defined as:
\begin{equation}
\label{eq.qu}
\begin{aligned}
& Q_g(x,y)=\sum_{i=1}^{n_v} Ch_i(x,y)\cos(2\psi_{gs}^i(x,y))\\
& U_g(x,y)=\sum_{i=1}^{n_v} Ch_i(x,y)\sin(2\psi_{gs}^i(x,y))\\
& \psi_g=\frac{1}{2}\tan^{-1}(\frac{U_g}{Q_g})
\end{aligned}
\end{equation}
The pseudo polarization angle $\psi_g$ is then defined correspondingly. Similar to the Planck polarization, $\psi_B=\psi_g+\pi/2$ gives the POS magnetic field orientation. The calculation for density/intensity gradient is in the same way and in this case we have $n_v=1$.

The relative alignment between magnetic fields orientation inferred from polarization $\phi_B$ and velocity gradient $\psi_B$ is quantified by the \textbf{Alignment Measure} (AM, \citealt{2017ApJ...835...41G}): 
\begin{align}
\label{AM_measure}
{\rm AM}=2(\langle \cos^{2} \theta_{r}\rangle-\frac{1}{2})
\end{align}
where $\theta_r=|\phi_B-\psi_B|$ and$\langle ...\rangle$ denotes the average within a region of interests. The value of AM spans from -1 to 1. AM = 1 mean $\phi_B$ and $\psi_B$ are parallel, while AM = -1 indicates $\phi_B$ and $\psi_B$ are perpendicular. The standard deviation divided by the sample size's square root gives the uncertainty $\sigma_{AM}$.

%%%%%%%%%%%%%%%%%%%%%%%%%%%%%%%%%%%%%%%%%%%%%%%%%%%%%%%%%%%%%%%%%%%%%%%%%%%%%%%%%%
\section{Results}
\label{sec:results}
\subsection{The central molecular zone}
\label{subsec:POS}
Fig.~\ref{fig:CMZ_B} presents the morphology of magnetic fields towards the CMZ measured by VGT. Here we use $\rm ^{12}CO$ and $\rm ^{13}CO$ emission lines. Pixels where the brightness temperature is less than three times the RMS noise level are blanked out. We average the gradients over each 20$\times$20 pixels sub-block, which is an empirically and numerically minimum block size \citep{LY18a}, and smooth the gradients map $\psi_g$ (see Eq.~\ref{eq.qu}) with a Gaussian filter $\rm FWHM\approx10'$. In what following, the usage of density/intensity gradient will be refereed as Gradient Technique (GT) and VGT specifies velocity gradient.

\begin{figure}
	\centering
	\includegraphics[width=1.0\linewidth]{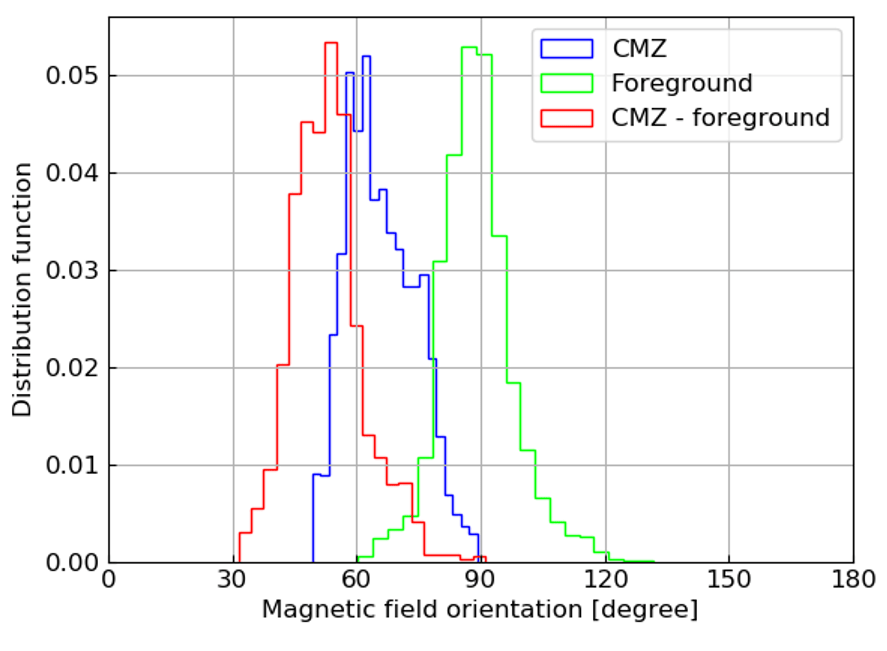}
	\caption{\label{fig:foregound} The histogram of magnetic field angle inferred from Planck 353 GHz polarization towards the CMZ and the galactic foreground. }
\end{figure}
\begin{figure}
	\centering
	\includegraphics[width=1.0\linewidth]{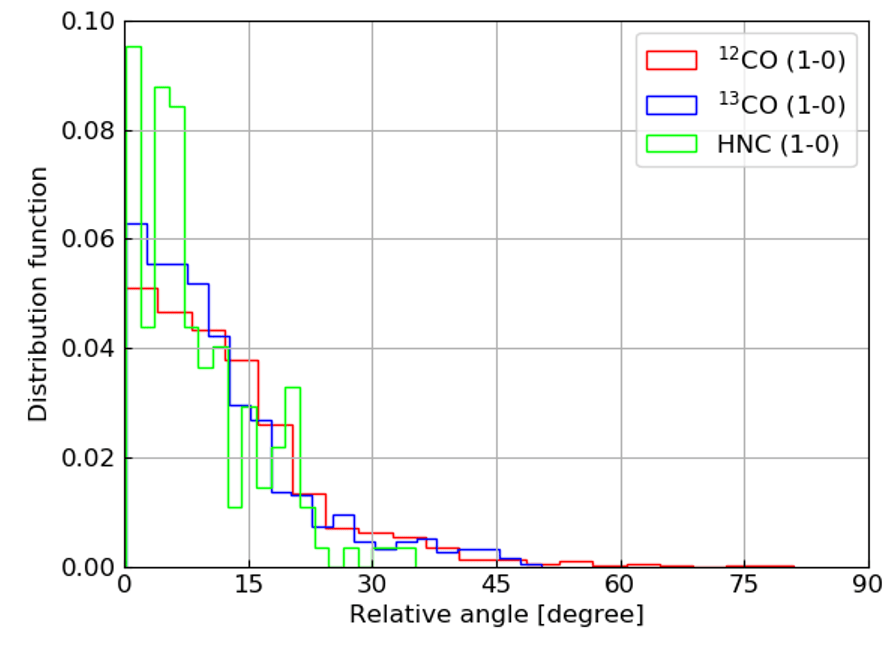}
	\caption{\label{fig:HRO} The histogram of relative angle between the magnetic field inferred from VGT and Planck 353 GHz polarization towards the CMZ.}
\end{figure}
\begin{figure*}
	\centering
	\includegraphics[width=1.0\linewidth]{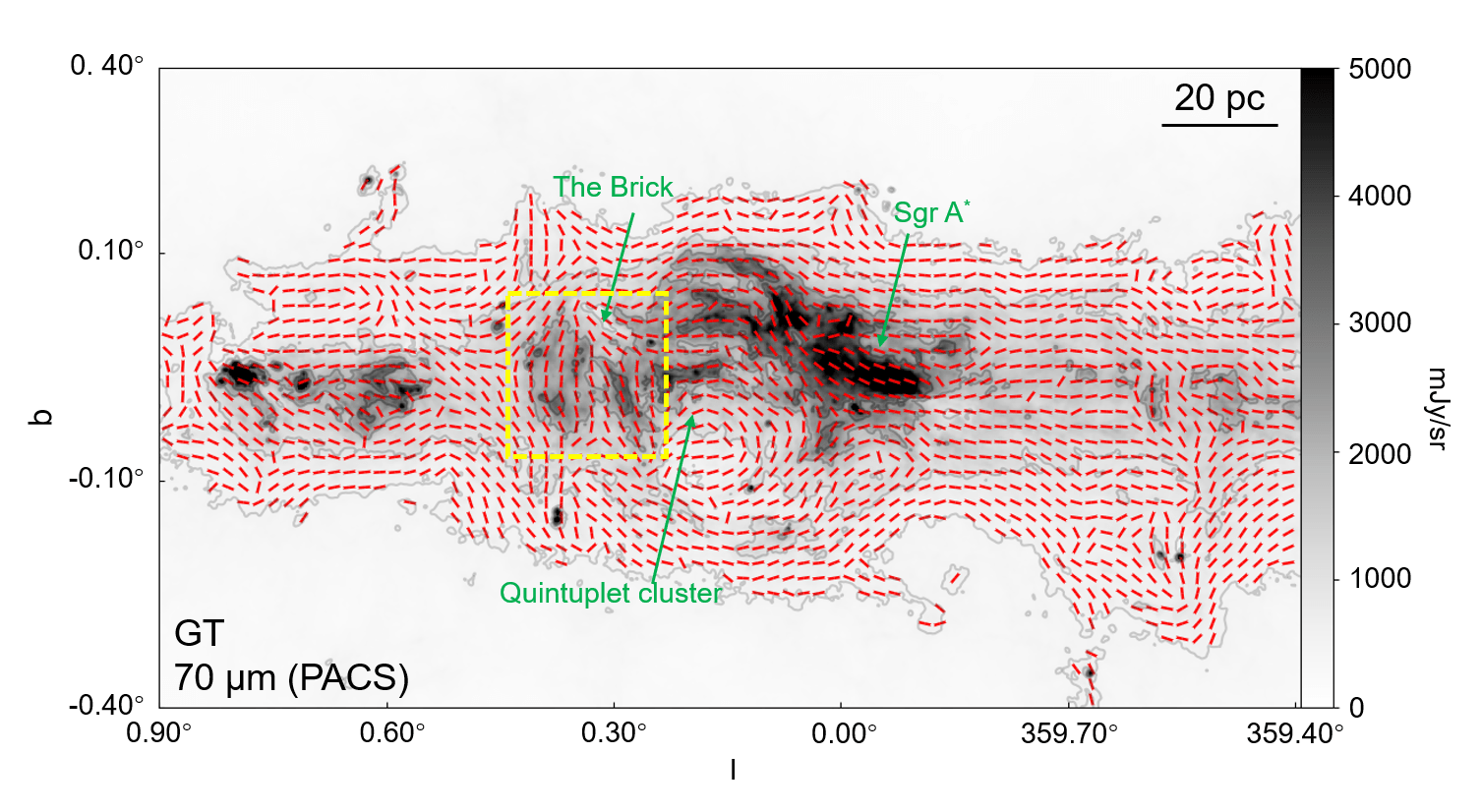}
	\caption{\label{fig:70um}The morphology of magnetic fields towards the CMZ region revealed by GT using the 70$\mu m$ PACS intensity image. The contours start from intensity value of 5000 mJy/sr. The yellow box outlines a region showing nearly vertical magnetic field.}
\end{figure*}

\begin{figure}
	\centering
	\includegraphics[width=1.0\linewidth]{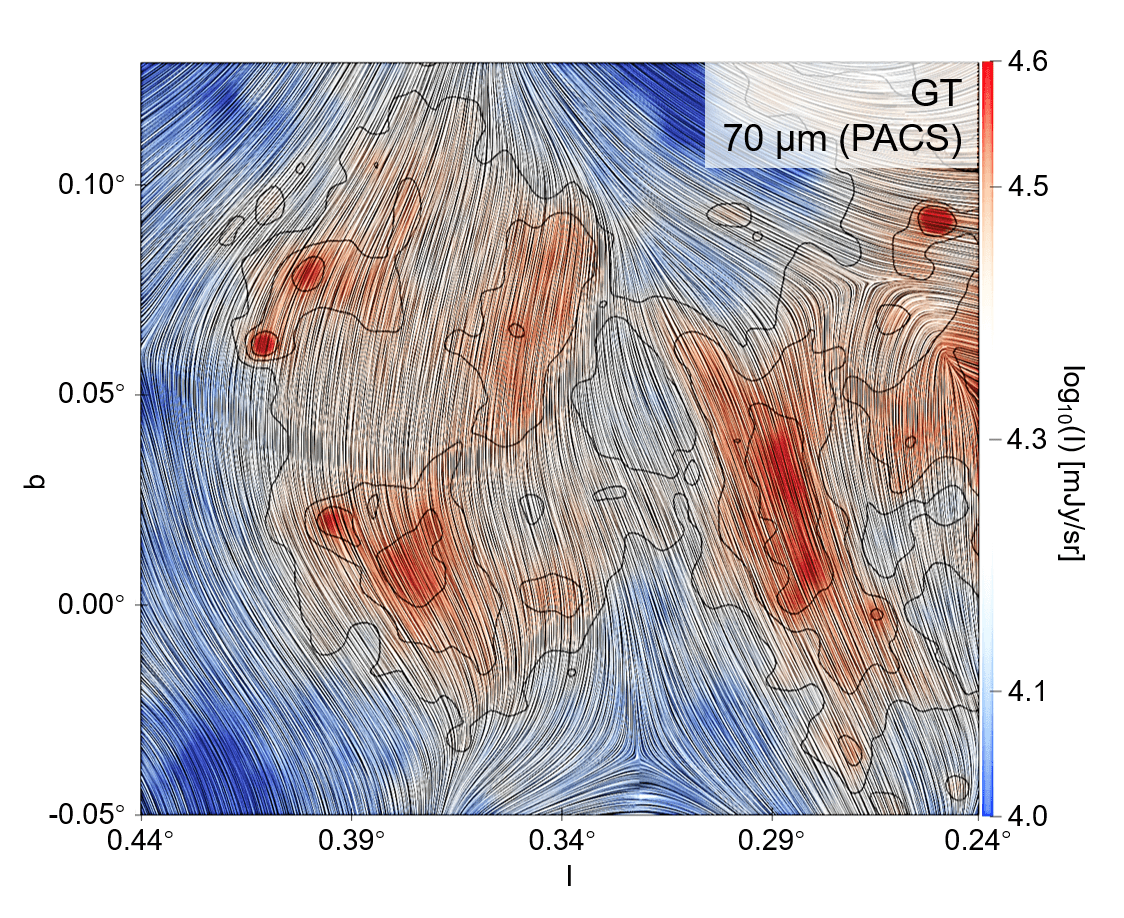}
	\caption{\label{fig:70um_zoom}The morphology of magnetic fields towards a zoom-in region highlighted by the yellow box in Fig.~\ref{fig:70um}. The contours start from intensity value of 20000 mJy/sr.}
\end{figure}

%\begin{figure*}
%\centering
%\includegraphics[width=1.0\linewidth,height=0.37\linewidth]{figures/RA_am.png}
%\caption{\label{fig:RA_am} \textbf{Left:} The map of the alignment measure towards the Radio Arc region. \textbf{Right:} The histogram of relative angle between the magnetic field inferred from VGT and HAWC+ polarization towards the Radio Arc region.}
%\end{figure*}
\begin{figure*}
	\centering
	\includegraphics[width=1.0\linewidth]{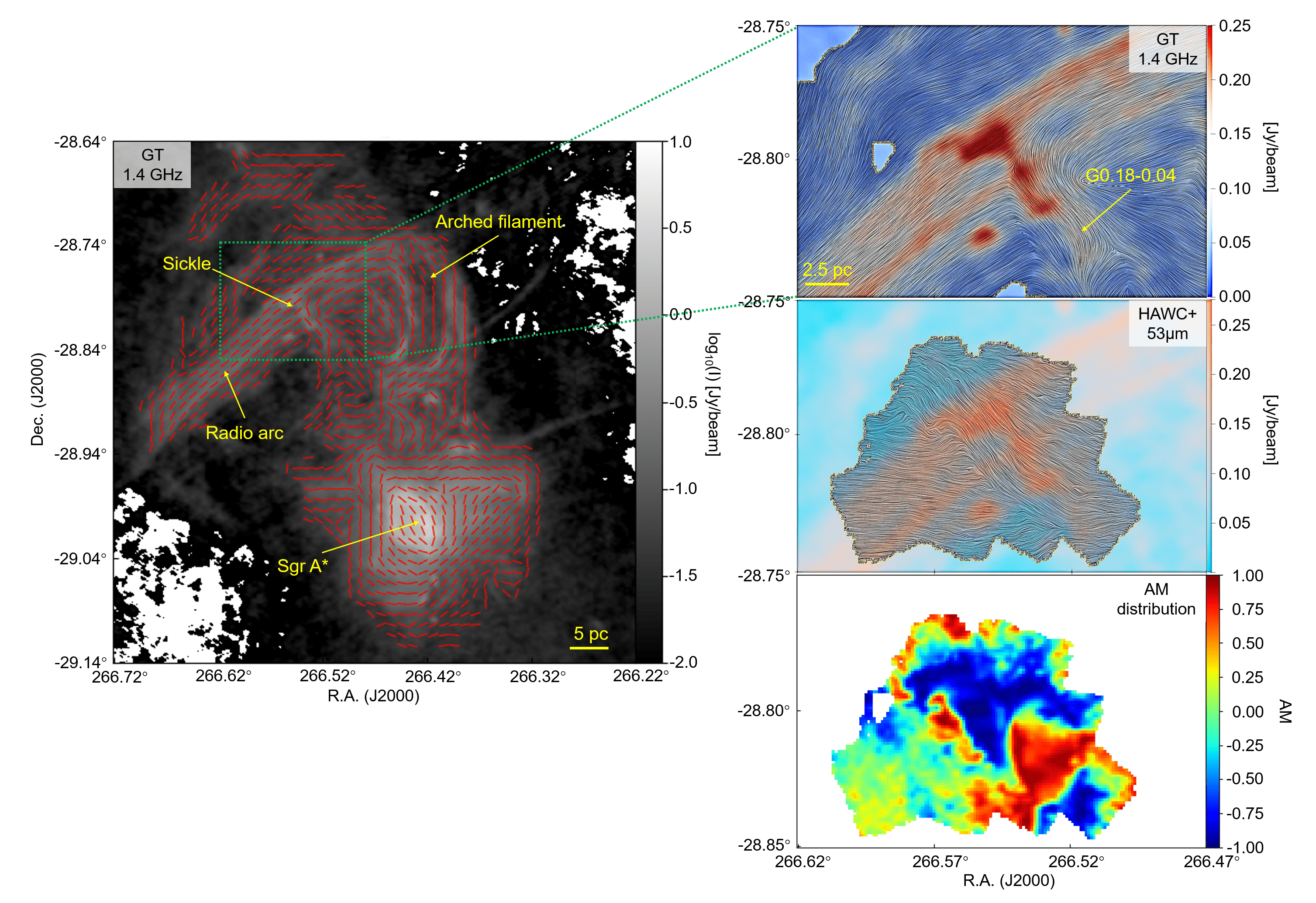}
	\caption{\label{fig:Radio_arc}\textbf{Left:} The morphology of magnetic fields towards the Radio Arc, Arched Filaments, and Sgr A east. The magnetic fields were inferred GT using the radio observation at 1.4 GHz. \textbf{Right:} The magnetic fields towards the Radio Arc (A zoom-in region corresponding to the dashed box in left panel). The magnetic fields were inferred from GT (top) and HAWC+'s measurement of polarized dust emission (middle). The distribution of AM (between GT and HAWC+) is given in the bottom panel.}
\end{figure*}

\begin{figure*}
	\centering
	\includegraphics[width=1.0\linewidth]{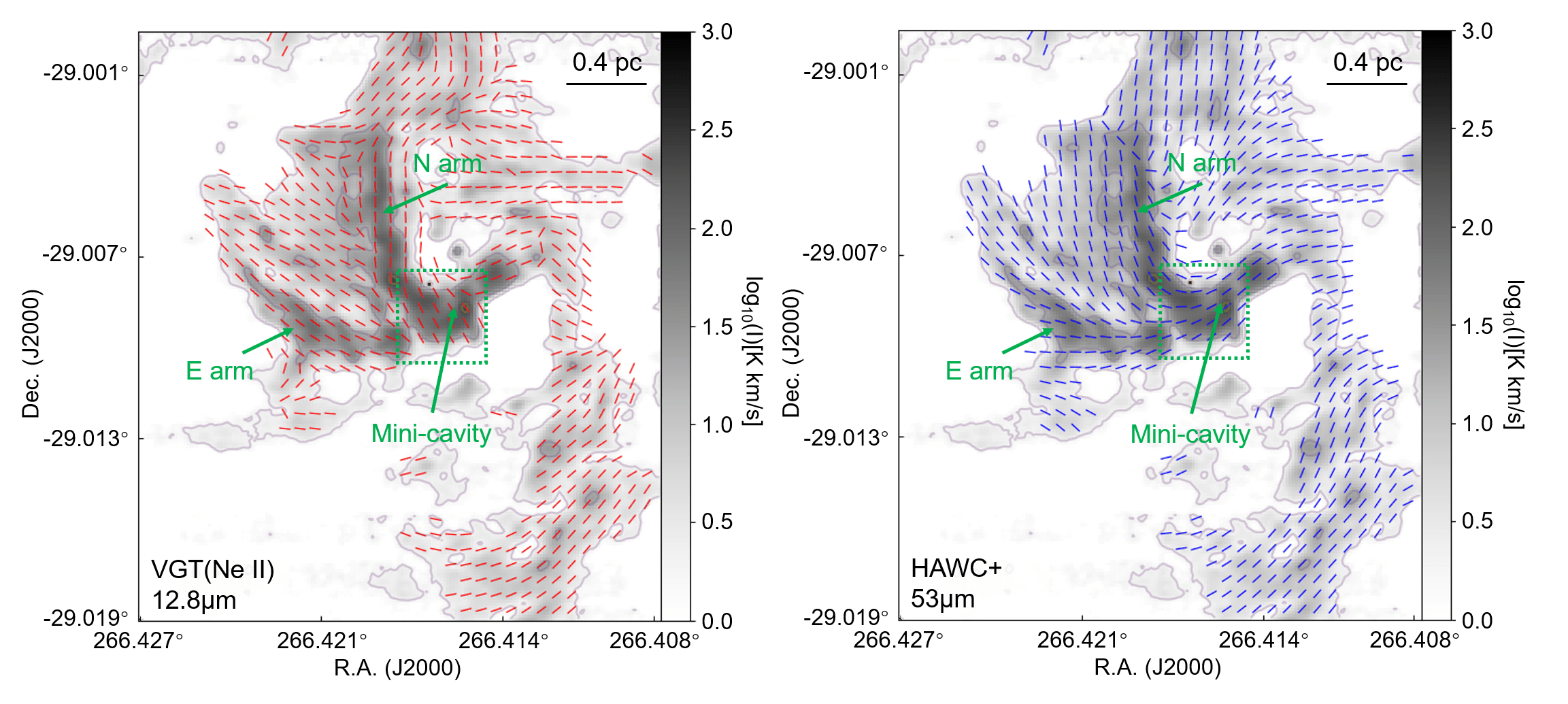}
	\caption{\label{fig:sgraA_ne} Differences and similarities of magnetic fields in different phases of gas towards the Sgr A West. {\bf Left}: morphology of magnetic fields towards Sgr A West region in ionized gas revealed by VGT using [Ne II] emission line. The magnetic field is overlaid with the integrated [Ne II] emission intensity map. {\bf Right}: The morphology of magnetic field, mostly in cold gas, revealed by HAWC+.}
\end{figure*}

We compare the magnetic field inferred from the Planck 353 GHz polarized dust signal data ($\rm FWHM\approx10'$).  In order to study the magnetic field morphology, it is necessary to isolate polarized dust emission originating in the CMZ from the diffuse polarized emission associated with Galactic foreground and background dust. To do so, we use two regions as the foreground reference point. One region is located at the west of the CMZ, spanning from $l = 3^\circ$ to 18$^\circ$ and $b = -0.35^\circ$ to $0.35^\circ$ and another one is located at the east of the CMZ spanning from $l = -18^\circ$ to $-3^\circ$ and $b = -0.35^\circ$ to $0.35^\circ$. By assuming that emission in these reference regions provides spatial uniformity foreground emission, we calculate the average Stokes $I$, $Q$, and $U$ in that region, and the mean values were then subtracted from each of the $I$, $Q$, $U$ maps of Planck polarization. Fig.~\ref{fig:foregound} presents the histograms of the magnetic field angle inferred from Planck polarization. The foreground region has a mean magnetic field $\approx90^\circ$ (in IAU convention) being parallel to the galactic plane. After the subtraction, the mean magnetic field shifts to $\approx50^\circ$ around. In what following, we use the foreground calibrated map for comparison. We re-grid the Planck polarization further to achieve the same pixel size as emission lines. 

We find the resulting gradients of $\rm ^{12}CO$ and $\rm ^{13}CO$ have good agreement with the Planck polarization showing $\rm AM = 0.85\pm0.01$ and $\rm AM = 0.88\pm0.01$, respectively. However, several apparent misalignment between the gradients of $\rm ^{12}CO$ and Planck polarization appear in the Sagittarius B region. It is likely that the dust polarization and the $\rm ^{12}CO$ emission probe different spatial regions. The optically thick tracer $\rm ^{12}CO$ samples the diffuse outskirt region of the cloud with volume density $n\approx10^2{\rm cm^{-3}}$, while dust polarization traces denser regions. The theory of Radiative Torque (RAT) alignment \citep{2007JQSRT.106..225L,2007ApJ...669L..77L} predicts that dust grains can remain aligned at high densities, especially in the presence of the embedded stars. Therefore, we expect that for $n\approx10^3{\rm cm^{-3}}$ grains are aligned in the regions that we study, as the misalignment disappears in $\rm ^{13}CO$'s gradients. This disagreement seen in $\rm ^{12}CO$ may also be caused by the bar-driven inflow hits the CMZ there. We repeat the analysis for denser gas tracer HNC. Similarly, we average gradients over each $20\times20$ pixels sub-block and smooth the gradient map with a Gaussian filter $\rm FWHM\approx10'$. The HNC's gradients result in a better alignment with the Planck polarization showing $\rm AM = 0.92\pm0.01$. In Fig.~\ref{fig:HRO}, we plot the histograms of the relative alignment between the VGT measurements and Planck polarization. The histograms appear as Gaussian distributions concentrating on the range of $0^\circ\sim10^\circ$. The median values of the histograms are approximately $10.67^\circ$, $8.78^\circ$, and $6.40^\circ$ for $\rm ^{12}CO$, $\rm ^{13}CO$, and HNC, respectively. As discussed above, the values may contain systematically angular differences since the VGT for a single molecular tracer contains only the information of magnetic fields within a certain gas density range.

We notice that in the east of the galaxy ($l\approx0.1^\circ$, $b\approx-0.15^\circ$), the gradients of $\rm ^{12}CO$ rapidly change their direction by $90^\circ$. This change may come from the superbubble around Quintuplet cluster. In this case of significant hydro effect, the velocity gradient is not expected to related with the magnetic field direction. Nevertheless, here we see that the VGT agrees the Planck polarization well globally, which indicates MHD turbulence is more critical in regulating the dynamics of the CMZ. 

Fig.~\ref{fig:70um} presents the magnetic field morphology inferred from the application of the GT to PACS's \SI{70}{\micro\meter} image. Similarly, we average the gradients over each 20$\times$20 pixels sub-block and blank out pixels where the intensity is less than 5000 mJy/sr. Finally, the gradient map is smoothed with a Gaussian filter $\rm FWHM\approx2'$. This measurement reveals the magnetic field to a different depth from the one measured by molecular lines (see Fig.~\ref{fig:CMZ_B}). Also note that in the case of shocks, the intensity gradient flips its direction by 90 degrees being parallel to the magnetic fields \citep{YL17b,IGs,H2}. Other physical processes, such as outflow, bubble, and \ion{H}{2} region, may also change the picture of MHD turbulence. Their effects require further study. In particular, we find agreement (between the molecular lines and \SI{70}{\micro\meter} image) in the Brick and Sgr A surroundings. However, disagreement also exists. In particular, the magnetic field morphology is more close to a divergence-free field in the Quintuplet cluster surroundings, at which the superbubble locates. This difference of magnetic fields inferred from different tracers is expected, as the molecular lines and \SI{70}{\micro\meter} image reveal the gas at a different LOS depth. Also, we zoom-in a small region near the Brick. As shown in Fig.~\ref{fig:70um_zoom}, the intensity structures are almost perpendicular to the galactic plane. Consequently, we find a vertical magnetic field there. This region has been called G0.33+0.04, G0.30+0.04, and G0.4+0.1 was suggested to be a supernova remnant superposed on another shell of nonthermal emission \citep{2000AJ....119..207L,1996MNRAS.283L..51K,2004ApJS..155..421Y,2015MNRAS.453..172P}.

\subsection{Radio Arc and Arched Filaments}
Fig.~\ref{fig:Radio_arc} presents the morphology of magnetic fields towards the Radio Arc and Arched Filaments measured by GT (i.e., intensity gradient) and HAWC+ polarized dust emission. For GT measurement, we blank out pixels where the intensity is less than three times the RMS noise level and average the gradients over each 20$\times$20 pixels sub-block. Finally, the resulting gradient map is smoothed with a Gaussian filter of  $\rm FWHM\approx1'$. Note here we are using the intensity gradient of synchrotron emission. Due  to  symmetry  of  velocity  and  magnetic  fluctuations in Alfv\'enic  turbulence \citep{2003MNRAS.345..325C},  the  gradients  of  magnetic  fluctuations are also perpendicular to local magnetic field. As synchrotron radiation arises from relativistic electrons spiraling along magnetic field lines, the anisotropic property is also entailed by the synchrotron emission \citep{2012ApJ...747....5L}. This gave rise to the developmentof the magnetic field tracing based on synchrotron intensity gradients (SIGs, see \citealt{2017ApJ...842...30L}).

The magnetic fields measured by GT follow the Radio Arc and the Arched Filaments. This measurement towards the Arched Filaments agrees with the results from the earlier HAWC+ polarization \citep{2003ApJ...599.1116C,2007SPIE.6678E..0DV}. The Radio Arc, which has been observed for decades, indicates a poloidal magnetic field perpendicular to the Galactic plane \citep{2019Natur.573..235H}. The GT measurement here first confirms this poloidal component. The recent polarization measurements from the Atacama Cosmology Telescope (ACT) suggests that the Radio Arc is invisible at frequencies above 224 GHz \citep{2021arXiv210505267G}. Nevertheless, the ACT measurement at 98 GHz shows the magnetic fields follow the Radio Arc, which confirms the results from GT. The elongation of radio structures along the magnetic field is expected to be the consequence of MHD turbulence as we discussed above. This is the result of the dynamics of motions induced by Alfv\'enic turbulence. Indeed, in the low Alfv\'en Mach numbers, $M_A$ the MHD turbulence is weak from the injection scale $L_{\rm inj}$ to the scale $L_{\rm inj}M_A^2$ \citep{LV99}. The cascade of weak turbulence does not change the parallel scale of perturbations but decreases their perpendicular scale. In other words, Alfv\'enic motions shred the medium into thinner and thinner filaments. Visually it may correspond to what is seen in \cite{2019ApJ...884..170P}, who found that the radio arc breaks up into parallel sub-filaments with magnetic fields directed along the filaments by using high-resolution radio images. These sub-filaments are not resolved in our GT measurement due to the averaging procedure. In this situation, the GT is detecting the large-scale intensity gradient of the ensemble of sub-filaments.

In the zoom-in Sickle region, the magnetic field inferred from HAWC+ polarization does not follow the Radio Arc but is parallel to the Galactic plane. In Fig.~\ref{fig:Radio_arc}, we plot the distribution of $\rm AM$ and the histogram of the relative angle, which reveals the anti-alignment ($\rm AM\approx-1$) of HAWC+ and the VGT towards the Radio Arc. This anti-alignment is likely coming from the fact that HAWC+ measures thermal emission while GT traces non-thermal components. In particular, in the G0.18-0.04 region and the Arched Filaments, which are not dominated by non-thermal emission, the GT measurement gives good agreement with the HAWC+ polarization.

%Note that unlike the cases of emission lines above, the \ion{H}{1} absorption, which weakens the background non-thermal emission at 21 cm, has little dynamical connection to the non-thermal filaments. Therefore, it is more likely that VGT's measurement is related to the morphology of the filaments. For instance, the morphology of the Radio Arc comes from their thermal emission, whereas the 21 cm absorption against the non-thermal emission filaments produces an apparent velocity gradient, which led to the VGT measurement of their morphology.
\begin{figure}
	\centering
	\includegraphics[width=1.0\linewidth]{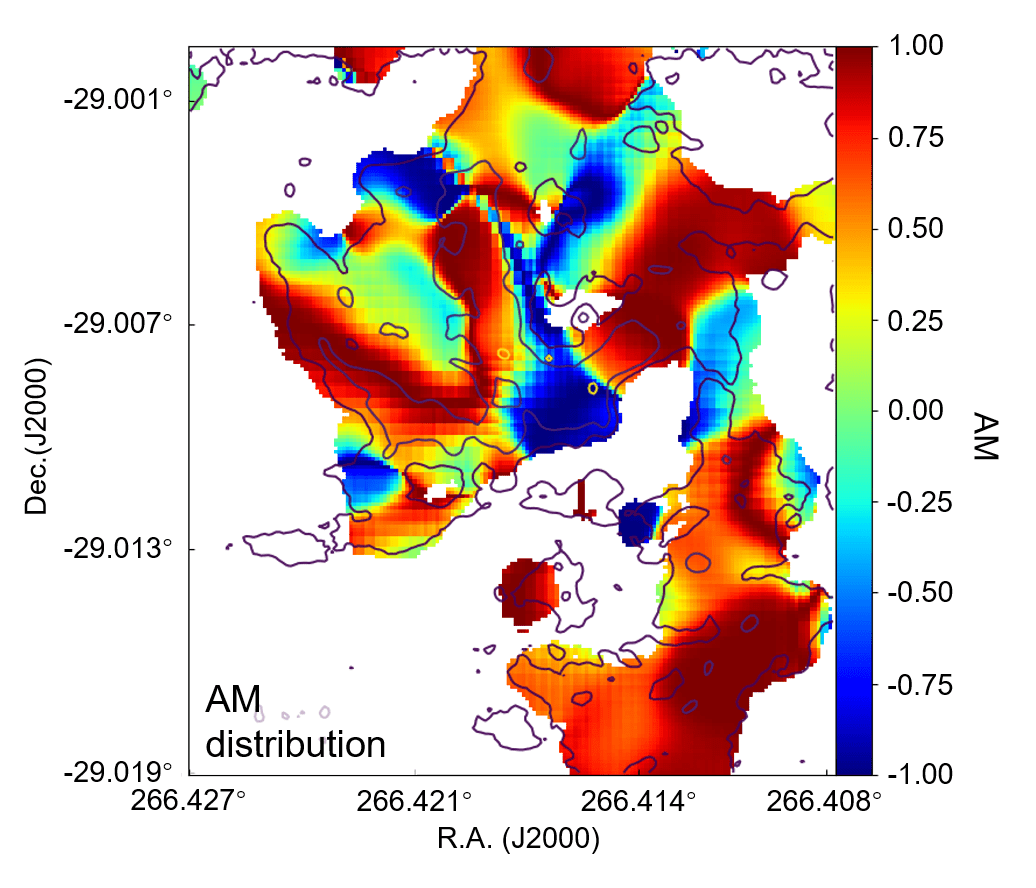}
	\caption{\label{fig:HRO_sgra} The distribution of alignment measurement (between GT and HAWC+) towards the Sgr A West region. The contours outline the structures seen in Fig.~\ref{fig:sgraA_ne}. }
\end{figure}
\begin{figure*}
	\centering
	\includegraphics[width=1.0\linewidth]{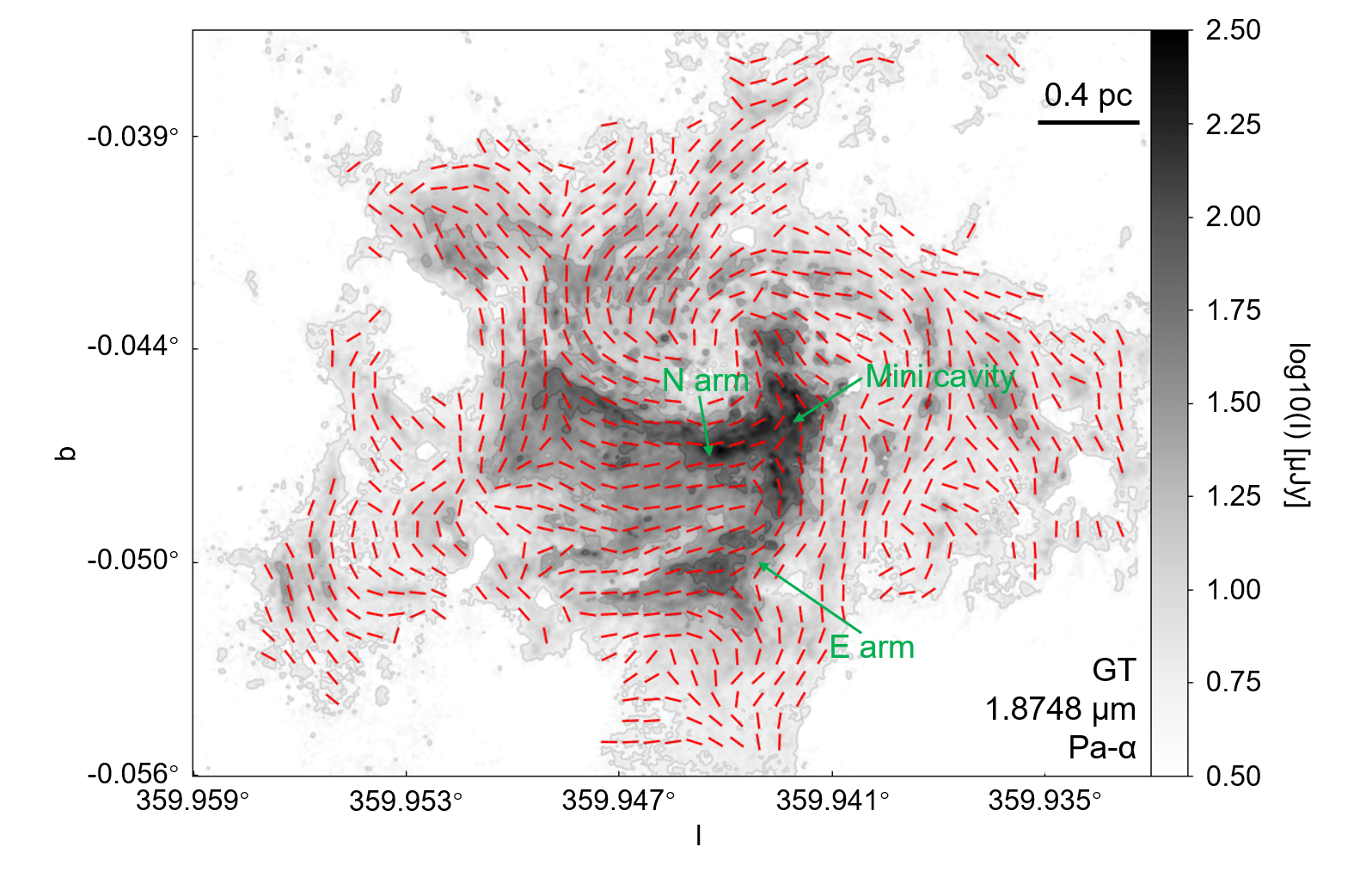}
	\caption{\label{fig:palpha} The morphology of magnetic fields towards the Sgr A west region in ionized gas revealed by GT using paschen-alpha emission image. The magnetic field is overlaid with the integrated paschen-alpha  emission intensity map. The contours start from intensity value of \SI{5}{\micro Jy}.}
\end{figure*}

\subsection{Sagittarius A west}
This section gives a close-up of the Sagittarius A* region and uses the same recipe for the VGT's calculation. Due to the high-resolution of the [Ne II] data, we smooth the gradient map with a Gaussian filter $\rm FWHM\approx5''$ to match the HAWC+ band A measurement. Pixels where the intensity is less than three times the RMS noise level are blanked out. Uncertainty maps are given in Appendix~\ref{appendix.a}.

In general, the magnetic fields traced by the VGT are globally compatible with the HAWC+ polarization measurements, although moderate discrepancy exists. As we discussed in \S~\ref{sec.dis}, a good agreement suggests turbulence's role is important. HAWC+ mostly samples magnetic fields in significantly cold gas, while [Ne II] samples ionized and diffuse gas. The regions, where magnetic fields in the two phases are aligned, suggest the magnetic fields' variation is small. Moreover, several small-scale structures exhibit prominent features. For instance,  the study of H30$\alpha$ mm hydrogen recombination line towards the same source in \cite{2019ApJ...872....2R} shows that the N arm has red-shifted radial velocity components $\approx100$ km s$^{-1}$, which are the velocities of peak emission. It gently decreases to $\approx0$ km s$^{-1}$ from north to south. Consequently, it makes a large velocity gradient that follows the N arm. However, we find a good correspondence of the VGT with HAWC+ polarization towards the N arm (see Fig.~\ref{fig:sgraA_ne}). Recall that the velocity gradient calculated via the VGT is rotated by 90 degrees here so that the VGT gradient is actually perpendicular to the N arm. As the VGT accounts for both turbulent velocity fluctuations and stream velocity, the agreement between HAWC+ and the VGT means that the contribution of turbulent velocity fluctuations dominates the velocity gradient. In addition, the E eastern arm's radial velocity components changes from red-shifted (at the east) to blue-shifted (near the mini cavity) \citep{2019ApJ...872....2R}. This change makes a velocity gradient following the E arm. However, the rotated VGT gradient still agrees with HAWC+, which means the MHD turbulence dominates the E arm and N arm.

As for the mini-cavity region (R.A.$\approx266.415^\circ$, Dec.$\approx-29.010^\circ$), the situation gets changed. The radial velocity components (i.e., the velocities of peak emission) induce a velocity gradient following the magnetic field, pointing from the east to the west \citep{2019ApJ...872....2R}. This velocity  gradient is on the scale of the entire stream and is thus a systematic one. However, in this region the (rotated) VGT anti-aligns with HAWC+ polarization (i.e, $\rm AM\approx-1$, see Figs.~\ref{fig:sgraA_ne} and \ref{fig:HRO_sgra}). This suggests that the contribution of ionized gas stream velocity dominates the velocity gradient, and the hydro effect is significant.

Furthermore, we give the magnetic field morphology associated with ionized Paschen-$\alpha$ gas in Fig.~\ref{fig:palpha}. We average the gradients over each 20$\times$20 pixels sub-block, blank out pixels where the intensity is less than 5 $\mu$Jy, and smooth the gradient map with a Gaussian filter $\rm FWHM\approx5''$. Note that here $\rm FWHM\approx5''$ is chosen to match HAWC+ measurements. A higher resolution map is flexible for GT's measurement. Similarly, we find that the resulting magnetic fields follow the N arm and E arm. In the mini cavity region, unlike the VGT's [Ne II] measurement, the magnetic field tends to be aligned with the HWAC+ polarization. As we discussed above, this difference may come from the fact that the velocity gradient in the mini-cavity region is dominated by stream velocity, which is not observable in the Paschen-$\alpha$ image. Again, Paschen-$\alpha$ samples ionized gas phase, while HAWC+ measures cold gas phase. 

%%%%%%%%%%%%%%%%%%%%%%%%%%%%%%%%%%%%%%%%%%%%%%%%%%%%%%%%%%%%%%%%%%%%%%%%%%%%%%%%%%
\section{Discussion}
\label{sec.dis}
\subsection{Tracing magnetic field with velocity and density gradients}
Tracing magnetic fields in the ISM is always difficult. Here we use the Gradient Technique (GT; \citealt{2017ApJ...835...41G,YL17a, LY18a, PCA} ) as an independent way to probe the magnetic fields. Unlike the typical polarization or Zeeman splitting measurements, GT employs either density gradient or velocity gradient to detect the anisotropy of MHD turbulence, which points to the local magnetic field direction.

As shown in this work, this novel technique GT can access magnetic fields of the ISM in various physical states. For instance, molecular species can be used to reveal the magnetic field in the cold gas phase. Employing multiple molecular emission lines, as shown in Fig.~\ref{fig:CMZ_B}, a 3D dimensional (i.e., the 3rd dimension is density) magnetic field tomography is achieved. In addition, due to the Doppler effect,  molecular emission lines contain the LOS velocity information. This velocity information could be used to separate the different molecular clouds (the galactic rotational curve is one possibility, see  \citealt{2019ApJ...874...25G}) or different molecular components of a single cloud. Together with multiple molecular emission lines, this would result in 4D magnetic field measurements.

This idea can also be implemented in dust emission images. Herschel provides the measurements at 70, 160, 250, 350, and \SI{500}{\micro\meter} \citep{2010A&A...518L...2P,2010A&A...518L...3G}. By employing the GT and the multi-wavelength data (including synchrotron emission and X-ray emission), 3D magnetic field maps are achievable across a much broader wavelength range than polarimetry. This brings a unique insight to separately study the magnetic fields embedded in multiple gas phases along the LOS to gauge their effects in different stages of astrophysical processes and at multiple length scales. 

\subsection{Contribution from systematic gradients}
Astrophysical objects may possess systematic gradients that are imposed by external conditions. The systematic gradients can contribute to our obtained velocity/density gradients. However, taking the entire CMZ as an example, the emission lines have beam resolution $\sim15''$ corresponding to $\sim0.5$ pc, at which the turbulence effect is already significant. The velocity gradient calculated per pixel must have the imprint of MHD turbulence’s anisotropy. The later averaging and smoothing processes enhance the anisotropy, as well as the correlation with respect to the magnetic fields. Secondly, at such a small scale ($\sim0.5$ pc), the velocity gradient contributed by galactic shear cannot be significant. As presented in Appendix~\ref{appendx:d}, the amplitude of turbulence's velocity gradient is more significant at small scales. We expect that at small scales turbulence's velocity gradient is dominant. Another contribution might come from the differential rotation of an object. It indeed contributes to the velocity gradient we obtained. However, this type of velocity gradient (from differential rotation) is not expected to have any correlation with the magnetic field. As here we see that the velocity gradient has good agreement with Planck polarization, the contribution from other effects, should not be significant.

The Sgr A west is a special region, in which the stream or other physical effects could be extremely strong. As shown in Appendix~\ref{appendx:c}, the Sgr A west has on average a LOS velocity dispersion $\sim 50 - 100$ km/s (see Fig.~\ref{fig:sgraA_dis}). This dispersion is contributed by turbulence, streams, outflows, and other factors, which are, however, indistinguishable. The dominance of turbulence could be accessed by comparing the velocity gradient with HAWC+ polarization because other factors are usually not expected to contribute to the correlation between the velocity gradient and the magnetic field. Therefore, a good agreement (of the velocity gradient and the magnetic field) suggests the importance of the MHD turbulence's role. However, in some particular situations, the correlation between the velocity gradient and magnetic field direction could also be produced by other mechanisms. For instance, the correlation can also appear due to the stretching of an initial “blobby” cloud into the current configuration by the supermassive black hole’s tidal field \citep{2018MNRAS.480.2939G}.

%%%%%%%%%%%%%%%%%%%%%%%%%%%%%%%%%%%%%%%%%%%%%%%%%%%%%%%%%%%%%%%%%%%%%%%%%%%%%%%%%%
\section{Summary} 
\label{sec:con}
We employ the GT to measure the magnetic field morphology in the CMZ in this work. Our analysis involves multiple-wavelength measurements, including the molecular emission and [Ne II] emission data cubes, as well as the radio 1.4 GHz continuum, \SI{70}{\micro\meter} Herschel image, and Paschen-alpha emission images. We compare the GT measured magnetic fields with Planck 353 GHz and HWAC+ polarized dust emissions. Our work provides a multi-scale, multi-phase, and polarimetry-independent view of the magnetism in various regions of the CMZ in differnt physical and chemical states. Our main discoveries are:
\begin{enumerate}
    \item The magnetic fields towards the entire CMZ traced by the VGT are globally compatible with the Planck polarization measurements in the large scale's order of 10 pc. This correspondence suggests that the magnetic field and turbulence are dynamically crucial in the galactic center. 
    \item By utilizing multiple molecular species, as well as \SI{70}{\micro\meter} Herschel image, we present the magnetic field morphology over various density ranges via the GT.
    \item The magnetic fields associated with the arched filaments and the thermal components of the radio arc agree with the HAWC+ polarization. 
    \item We present the magnetic field morphology towards the radio arc. The magnetic fields associated with the non-thermal radio arc are perpendicular to the HAWC+ polarization and the galactic plane, revealing the poloidal magnetic field components in the galactic center.
    \item We find the magnetic fields associated with ionized gas phase in the Sgr A West region, which are traced [Ne II] emission line via the VGT, are in agreements with the one measured by the HAWC+ polarization, which traces the cold dust phase. However, in several distinct special regions, for instance, the mini-cavity, the VGT measurement is perpendicular to HAWC+ polarization.
    \item We apply GT to predict the magnetic field morphology towards the Sgr A region, using the Paschen-alpha emission image down to scale of 0.2 pc.
\end{enumerate}
 
\section*{Acknowledgements}
Y.H. acknowledges the support of the NASA TCAN 144AAG1967. A.L. acknowledges the support of the NSF grant AST 1816234 and NASA ATP AAH7546. Based Sgr A and Sickle on observations made with the NASA/DLR Stratospheric Observatory for Infrared Astronomy (SOFIA). SOFIA is jointly operated by the Universities Space Research Association, Inc. (USRA), under NASA contract NNA17BF53C, and the Deutsches SOFIA Institut (DSI) under DLR contract 50 OK 0901 to the University of Stuttgart. We acknowledge the Nobeyama Observatory for providing the data of the CMZ region.
%\software{Julia \citep{2012arXiv1209.5145B}, ZEUS-MP/3D \citep{2006ApJS..165..188H}, SPARX\citep{2019ApJ...873...16H}, Paraview \citep{ayachit2015paraview}}
\section*{Data availability}
The data underlying this article will be shared on reasonable request to the corresponding author.

%\newpage
%\bibliography{sample63}{}
%\bibliographystyle{aasjournal}

\appendix
\section{Uncertainty of the magnetic field direction measured by GT}
\label{appendix.a}
\begin{figure*}
\centering
\includegraphics[width=1.0\linewidth]{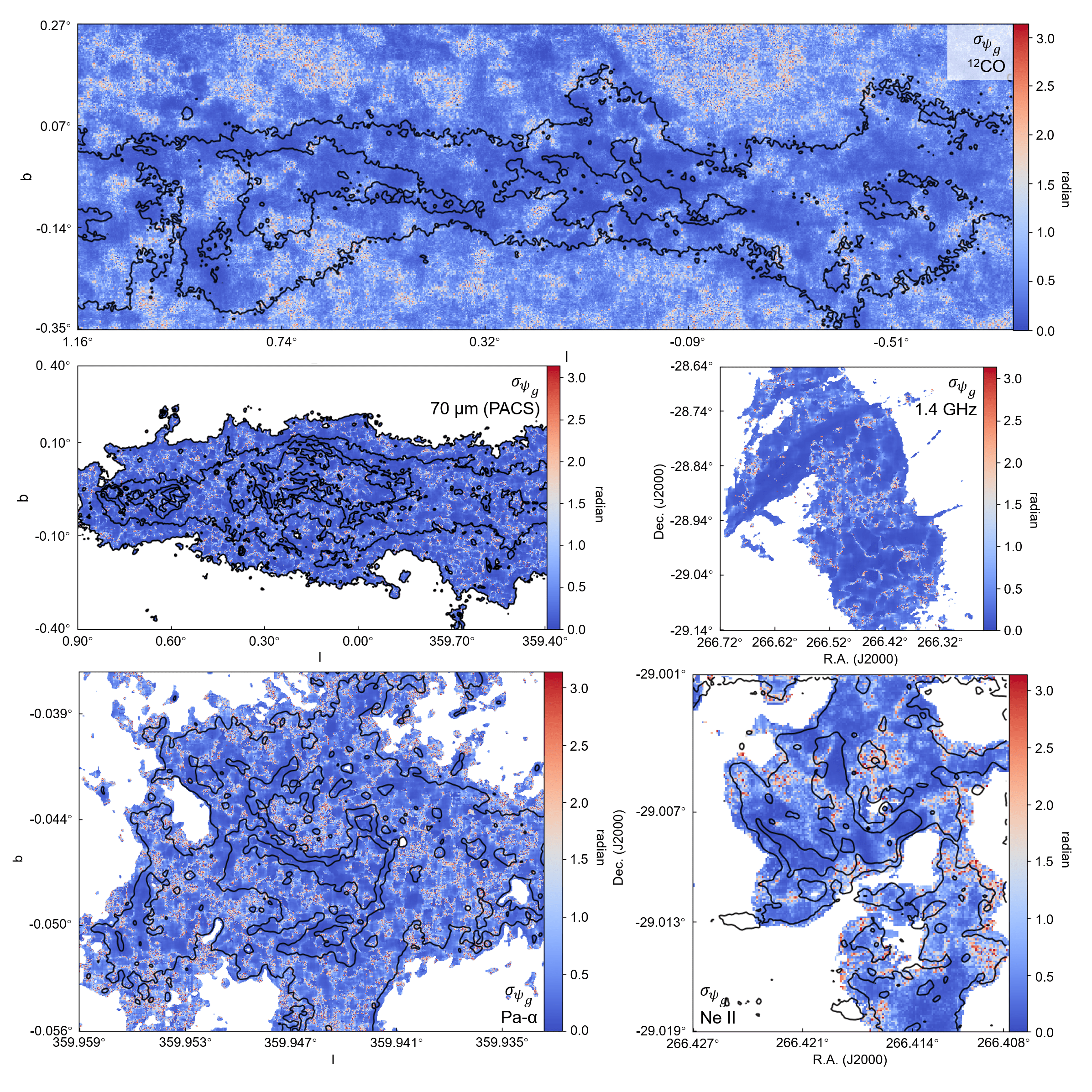}
\caption{\label{fig:error} Uncertainty maps for the magnetic field measured by GT or VGT.}
\end{figure*}

The two significant uncertainties of the magnetic field can come from the systematic error in the observational map and the GT algorithm. Recall that GT takes a subregion and fits a corresponding Gaussian histogram of the gradient’s orientation. The output of GT only takes the statistically most crucial angle, i.e., the angle of orientation corresponding to the Gaussian fitting peak value of the histogram. This procedure incidentally suppresses the part of the systematic noise in the observation map. The uncertainty in every single pixel can be considered as the error $\sigma_{\psi_{gs}}(x,y,v)$ from the Gaussian fitting algorithm within the 95\% confidence level.

Considering the noise $\sigma_n(x,y,v)$ in velocity channel $Ch(x,y,v)$ and error propagation, the uncertainties $\sigma_{Q}(x,y)$ and $\sigma_{U}(x,y)$ of the Pseudo Stokes parameters $Q_g(x,y)$ and $U_g(x,y)$ can be obtained from:
\begin{equation}
    \begin{aligned}
    \sigma_{\cos}(x,y,v)&=|2\sin(2\psi_{gs}(x,y,v))\sigma_{\psi_{gs}}(x,y,v)|\\
    \sigma_{\sin}(x,y,v)&=|2\cos(2\psi_{gs}(x,y,v))\sigma_{\psi_{gs}}(x,y,v)|\\
    \sigma_{q}(x,y,v)&=|Ch\cdot\cos(2\psi_{gs})|\sqrt{(\sigma_n/Ch)^2+(\sigma_{\cos}/\cos(2\psi_g))^2}\\
    \sigma_{u}(x,y,v)&=|Ch\cdot\sin(2\psi_{gs})|\sqrt{(\sigma_n/Ch)^2+(\sigma_{\sin}/\sin(2\psi_g))^2}\\
    \sigma_{Q}(x,y)&=\sqrt{\sum_{v}\sigma_{q}(x,y,v)^2}\\
    \sigma_{U}(x,y)&=\sqrt{\sum_{v}\sigma_{u}(x,y,v)^2}\\
    \sigma_{\psi_{g}}(x,y)&=\frac{|U_g/Q_g|\sqrt{(\sigma_{Q}/Q_g)^2+(\sigma_{U}/U_g)^2}}{2[1+(U_g/Q_g)^2]}
    \end{aligned}
\end{equation}
where $\sigma_{\psi_{g}}(x,y)$ gives the angular uncertainty of the resulting magnetic field direction. For single intensity image, the uncertainty is defined in a similar way. In this case, we have $\sigma_{Q}(x,y)=\sigma_{q}$ and $\sigma_{U}(x,y)=\sigma_{u}$. The uncertainty maps are presented in Fig.~\ref{fig:error}. The $\rm ^{12}CO$ map exhibits more significant uncertainty due to the accumulation along the LOS. Nevertheless, the median value for all maps is less than $15^\circ$.

\section{Foreground and background correction}
\label{appendx:b}
\begin{figure*}
\centering
\includegraphics[width=1.0\linewidth]{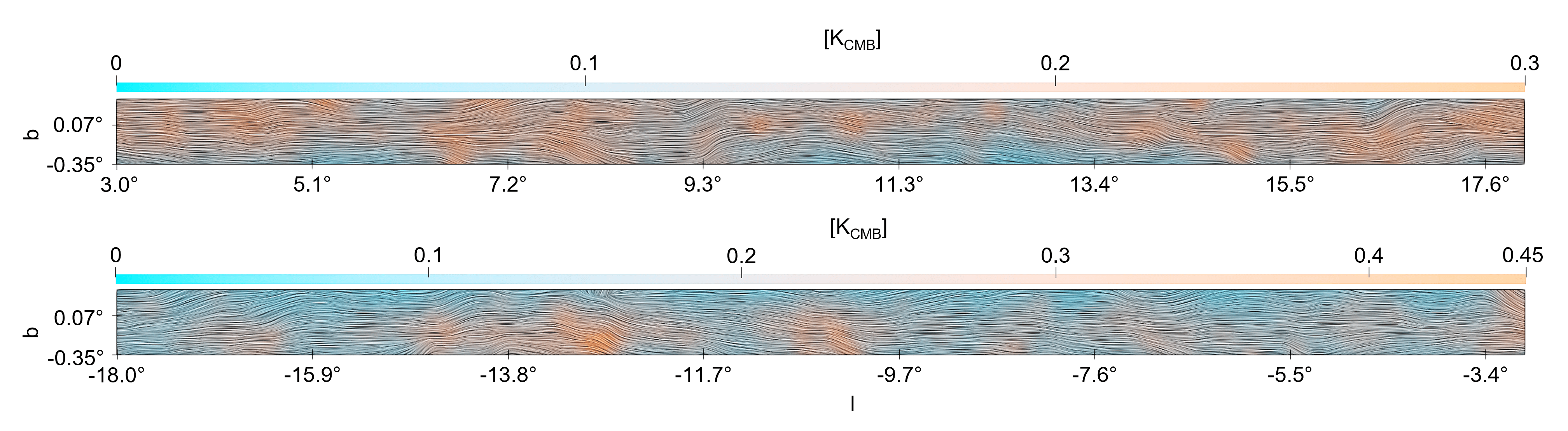}
\caption{\label{fig:fore_map} Visualization of magnetic fields towards the foreground regions used for correction. The magnetic fields were inferred from Planck 353 GHz polarized dust emission.}
\end{figure*}

In Fig.~\ref{fig:fore_map}, we show the two regions used for the foreground and background correction. One region is located at the west of the CMZ, spanning from $l = 3^\circ$ to 18$^\circ$ and $b = -0.35^\circ$ to $0.35^\circ$ and another one is located at the east of the CMZ spanning from $l = -18^\circ$ to $-3^\circ$ and $b = -0.35^\circ$ to $0.35^\circ$. In generally the magnetic field orientation is horizontal being along the galactic disk. 

To correction the foreground/background's contribution, we compute the averaged values of Stokes parameters $\bar{Q}$ and $\bar{U}$ over the two regions. Then we subtract $\bar{Q}$ and $\bar{U}$ from the Stokes parameters $Q$ and $U$ of the CMZ region:
\begin{equation}
\begin{aligned}
    U'&=U-\bar{U}\\
    Q'&=Q-\bar{Q}
\end{aligned}
\end{equation}

The corrected polarization angle is then:
\begin{equation}
    \phi'=\frac{1}{2}\arctan(-U', Q')
\end{equation}

\section{Velocity dispersion of the Sgr A west region}
\label{appendx:c}
\begin{figure*}
\centering
\includegraphics[width=1.0\linewidth]{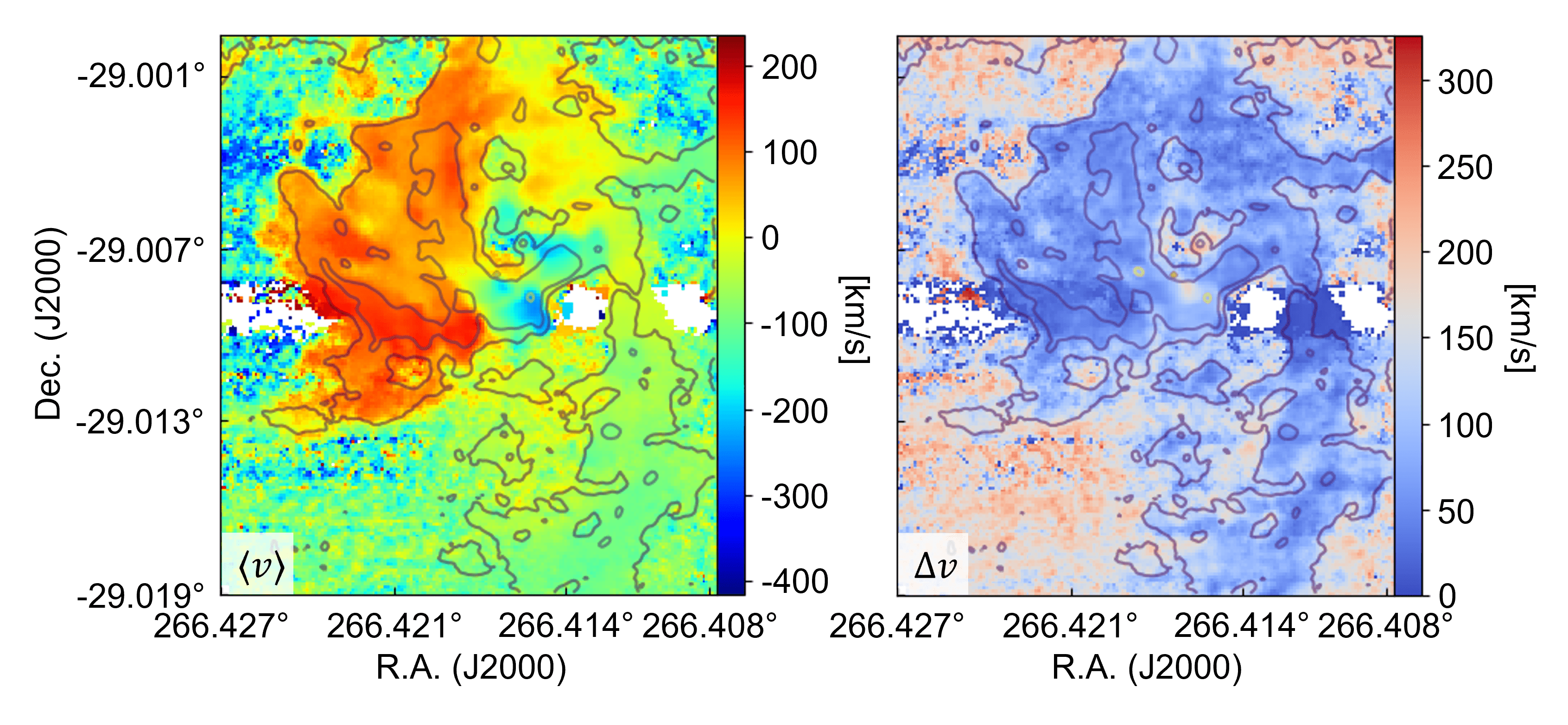}
\caption{\label{fig:sgraA_dis} Moment maps $\langle v\rangle$ (left) and $\Delta v$ (right) for the [Ne II] line.}
\end{figure*}
We present higher order moments to study the velocity structure of the Sgr A west region. The first moment map gives the intensity weighted and averaged LOS velocity $\langle v\rangle$:
\begin{equation}
    \langle v\rangle=\frac{\int Ch(x,y,v) v dv}{\int Ch(x,y,v)  dv}
\end{equation}
where $Ch(x,y,v)$ is the channel map's intensity. The second moment map gives the LOS velocity dispersion $\Delta v$:
\begin{equation}
    \Delta v= \sqrt{\frac{\int Ch(x,y,v)(v-\langle v\rangle)^2dv}{\int Ch(x,y,v)  dv}}
\end{equation}

As shown in Fig.~\ref{fig:sgraA_dis}, the $\langle v\rangle$ maps exhibits a significantly higher velocity $\sim200$ km/s in the north part compared to the rest of the region. The $\Delta v$ map within the [Ne II] contours (see Fig.~\ref{fig:sgraA_ne}) show that the Sgr A west has on average a LOS velocity dispersion $\sim 50 - 100$  km/s. This dispersion is contributed by turbulence,  streams, and other factors.

\section{Gradient amplitude}
\label{appendx:d}
\begin{figure*}
\centering
\includegraphics[width=0.45\linewidth]{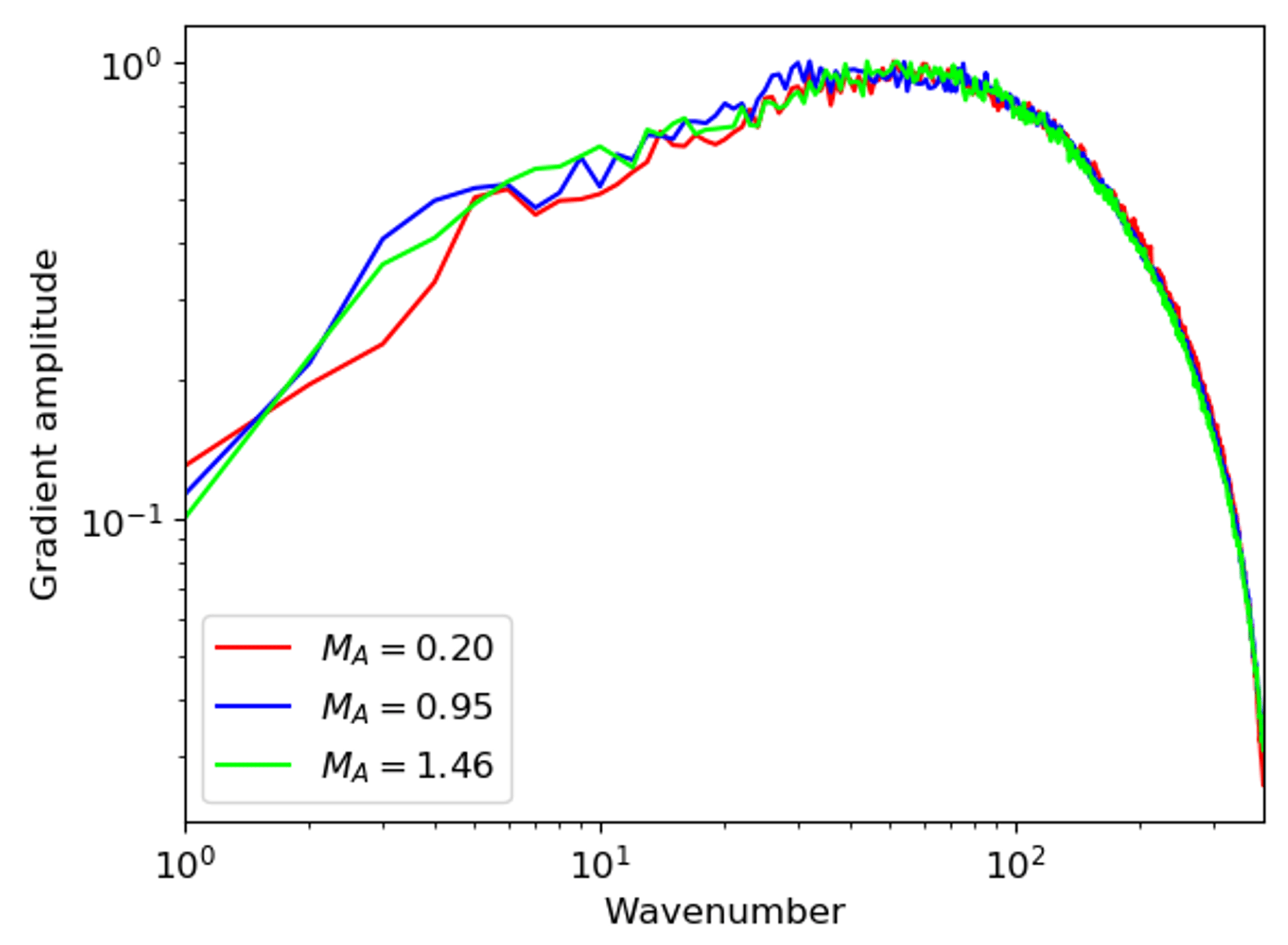}
\caption{\label{fig:run}Normalized (velocity channel) gradient amplitude as a function of wavenumber. All simulations have $M_S\sim5.0$.}
\end{figure*}

We numerically explore the amplitude of velocity gradient at different spatial scales. The numerical 3D MHD simulations
are generated by the ZEUS-MP/3D code \citep{2006ApJS..165..188H}, which simulates a single-fluid and operator-split MHD turbulence in Eulerian frame. Periodic boundary conditions and solenoidal turbulence injections (at wavenumber $k=2$) are applied in our simulations, which are regularly grid into 792$^3$ cells. Details of numerical setup can be found in \cite{H2}.

Here we use three supersonic MHD simulations, which have the sonic Mach number $M_S\sim5.0$. The simulations also corresponds to sub-Alfv\'enic ($M_A\sim0.20$), trans-Alfv\'enic ($M_A\sim0.95$), and super-Alfv\'enic ($M_A\sim1.46$), respectively. we generate three synthetic spectroscopic PPV cubes from the simulations and calculate the amplitude of thin velocity channel gradients:
\begin{equation}
|\nabla Ch(x,y)|=\sqrt{|\nabla_x Ch(x,y)|^2+|\nabla_y Ch(x,y)|^2}
\end{equation}
when $Ch(x,y)$ represents central channel, which corresponds to the averaged emission line maximum and satisfies the thin channel criterion (see Eq.~\ref{eq1}). Also, to study the amplitude of gradient at different spatial scale, we perform Fourier transform to the channel map and remove spatial frequencies of interests. For the sake of simplicity, here we remove $n>k>n-1$, where $n$ is an integer, and  apply the inverse Fourier transform to the processed map and then perform the gradients' calculation.  

Fig.~\ref{fig:run} shows the normalized and averaged gradient amplitude as a function of wavenumber. The value of the x-axis means only the corresponding wavenumber is left. We see that the amplitude increases with the increment of wavenumber until $k\sim80$, which is the dissipation scale. Note that large wavenumber corresponds to small spatial scale. Therefore, the results suggest that the gradients at small spatial scales have a dominant role.

%% This command is needed to show the entire author+affiliation list when
%% the collaboration and author truncation commands are used.  It has to
%% go at the end of the manuscript.
%\allauthors

%% Include this line if you are using the \added, \replaced, \deleted
%% commands to see a summary list of all changes at the end of the article.
%\listofchanges
% Don't change these lines
\bsp	% typesetting comment
\label{lastpage}
\end{document}